\documentclass[aps,prl,twocolumn,groupedaddress]{revtex4}

\usepackage{graphicx}
\usepackage{dcolumn}
\usepackage{bm}
\usepackage{subfigure}
\usepackage{amsfonts}
\usepackage{amssymb,epsf}

\begin{document}

\title{Charged BTZ-like Black Holes in Higher Dimensions}
\author{S. H. Hendi\footnote{email address: hendi@mail.yu.ac.ir}}
\affiliation{Physics Department, College of Sciences, Yasouj University, Yasouj 75914,
Iran\\
Research Institute for Astrophysics and Astronomy of Maragha
(RIAAM), P.O. Box 55134-441, Maragha, Iran}

\begin{abstract}
Motivated by many worthwhile paper about $(2+1)$-dimensional BTZ
black holes, we generalize them to to $(n+1)$-dimensional
solutions, so called BTZ-like solutions. We show that the electric
field of BTZ-like solutions is the same as $(2+1)$-dimensional BTZ
black holes, and also their lapse functions are approximately the
same, too. By these similarities, it is also interesting to
investigate the geometric and thermodynamics properties of the
BTZ-like solutions. We find that, depending on the metric
parameters, the BTZ-like solutions may be interpreted as black
hole solutions with inner (Cauchy) and outer (event) horizons, an
extreme black hole or naked singularity. Then, we calculate
thermodynamics quantities and conserved quantities, and show that
they satisfy the first law of thermodynamics. Finally, we perform
a stability analysis in the canonical ensemble and show that the
BTZ-like solutions are stable in the whole phase space.
\end{abstract}

\maketitle

The discovery and investigation of the $(2+1)$-dimensional BTZ
(Banados-Teitelboim-Zanelli) black holes \cite{BTZ1,BTZ2,BTZ3}
organizes one of the great advances in gravity because the provide
a simplified model for exploring some conceptual issues, not only
about realization of the black hole thermodynamics
\cite{Carlip95,Ashtekar02,Sarkar06} but also about developments in
quantum gravity, string and gauge theory, and specially in the
context of the AdS/CFT conjecture \cite{Witten98,Carlip05}.

The BTZ black hole is suitable, in a nice way, for the AdS/CFT framework and
perform a central role in recent investigations and improve our
comprehension of low dimensional gravity and of general feature of the
gravitational interaction \cite{Witten07}.

Interest in the BTZ black hole has recently heightened with the
discovery that the thermodynamics of higher-dimensional black
holes (see for e.g. \cite{Thermodynamics,PMIpaper,Conformalpaper})
can often be understood in terms of the BTZ solution. The entropy
of the BTZ black hole as we consider it here, grown from the
papers \cite{Saida99,Cadoni08} in the case of AdS3 (and also see
\cite{Larranaga10}). Some class of higher dimensional black holes
contains the BTZ black holes in the near-horizon region have been
studied in \cite{Hyun97,Sfetsos98}. Also, BTZ-like black holes in
even dimensional Lovelock gravity has been investigated in
\cite{Canfora10} and some solutions of the BTZ black hole in every
dimension by defining the singularity as the closed orbits have
been discussed in \cite{Claessens09}.

In recent years there has been increasing interest about black hole
solutions whose matter source is power Maxwell invariant, i.e., $\left(
F_{\mu \nu }F^{\mu \nu }\right)^{s}$ \cite{PMIpaper}. This theory is
considerably richer than that of the linear electromagnetic field and in the
special case ($s=1$) it can reduces to linear field. In addition, in $(n+1)$
-dimensional gravity, for the special choice $s=(n+1)/4$, matter source
yields a traceless Maxwell's energy-momentum tensor which leads to conformal
invariance. The idea is to take advantage of the conformal symmetry to
construct the analogues of the four dimensional Reissner-Nordstr\"{o}m
solutions in higher dimensions \cite{Conformalpaper}. Also, it is valuable
to find and analyze the effects of exponent $s$ on the behavior of the new
solutions, when $s=n/2$. In this case the solutions are completely different
from another cases ($s\neq n/2$).

One of the important defects of higher dimensional solutions in
Einstein-Maxwell gravity is that in 3-dimension, these solutions
does not reduce to BTZ black hole. The aim of this paper is to
consider a class of nonlinear electrodynamics field coupled to
Einstein gravity and introduce higher dimensional charged BTZ-like
black hole. In this solution, the lapse function has a logarithmic
term and also the electromagnetic field is proportional to
$r^{-1}$.

\subsection{Field Equations of Einstein Gravity with nonlinear Electromagnetic Source \label{Fiel}}
The $(n+1)$-dimensional action in which gravity is coupled to
nonlinear electrodynamics field is given by
\begin{equation}
\mathcal{I}=-\frac{1}{16\pi
}\int\limits_{\mathcal{M}}d^{n+1}x\sqrt{-g} \left(
\mathcal{R}-2\Lambda -\left( \alpha \mathcal{F}\right) ^{s}\right)
, \label{Action}
\end{equation}
where $\mathcal{R}$ is scalar curvature, $\Lambda $ refers to the
negative cosmological constant which is in general equal to
$-n(n-1)/2l^{2}$ for asymptotically AdS solutions, in which $l$\
is a scale length factor, $\alpha $ is a constant in which we
should fix it and $s$ is the power of nonlinearity and hereafter
we set it to $n/2$ to attain BTZ-like solutions. Varying the
action (\ref{Action}) with respect to the metric $g_{\mu \nu }$
and the gauge field $A_{\mu }$, (with $s=n/2$) the field equations
are obtained as
\begin{equation}
R_{\mu \nu }-\frac{1}{2}g_{\mu \nu }(\mathcal{R}-2\Lambda )=\alpha \left(
\alpha \mathcal{F}\right) ^{n/2-1}\left( \frac{1}{2}g_{\mu \nu }\mathcal{F-}
nF_{\mu \lambda }F_{\nu }^{\;\lambda }\right) ,  \label{GravEq}
\end{equation}
\begin{equation}
\partial _{\mu }\left( \sqrt{-g}F^{\mu \nu }\left( \alpha \mathcal{F}\right)
^{n/2-1}\right) =0.  \label{MaxEq}
\end{equation}
Now, we should fix the sign of the constant $\alpha $ in order to ensure the
real solutions. It is easy to show that for static diagonal metric in which
the nonzero component of $A_{\mu }$ is $A_{0}$, we have
\[
\mathcal{F}=F_{\mu \nu }F^{\mu \nu }=-2\left( \frac{dA_{0}}{dr}\right) ^{2},
\]
and so the power Maxwell invariant, $\left( \alpha
\mathcal{F}\right) ^{n/2}$, may be imaginary for positive $\alpha
$, when $n/2$ is fractional (for even dimension). Therefore, we
set $\alpha =-1$, to have real solutions without loss of
generality.

\subsection{(2+1)-dimensional charged BTZ solution \label{BTZ}}
The field equations of three dimensional solution is the same as
equations (\ref{GravEq}) and (\ref{MaxEq}), when we set $n=2$. The
charged BTZ black hole is a solution of the $(2+1)$-dimensional
Einstein-Maxwell gravity with a negative cosmological constant
$\Lambda =-1/l^{2}$. The metric is given by \cite{BTZ1}
\begin{equation}
ds^{2}=-N^{2}(r)dt^{2}+\frac{dr^{2}}{N^{2}(r)}+r^{2}d\varphi ^{2},
\label{BTZmetric}
\end{equation}
Here, we use the gauge potential ansatz
\begin{equation}
A_{\mu }=h(r)\delta _{\mu }^{0}  \label{Amu1}
\end{equation}
in the electromagnetic field equation (\ref{MaxEq}). We obtain
\begin{equation}
rh^{\prime \prime }(r)+h^{\prime }(r)=0,  \label{heq}
\end{equation}
where prime and double prime are first and second derivative with
respect to $r$ , respectively. One can show that the solution of
Eq. (\ref{heq}) is $h(r)=q\ln (\frac{r}{l})$ and the electric
field in $(2+1)$-dimension is given by
\begin{equation}
F_{tr}=\frac{q}{r}.  \label{Ftr}
\end{equation}
To find the metric function of (\ref{BTZmetric}), one may use any
components of Eq. (\ref{GravEq}) for $n=2$. Considering the
function $h(r)$, the solution of Eq. (\ref{GravEq}) can be written
as
\begin{equation}
N^{2}(r)=\frac{r^{2}}{l^{2}}-\left[ M+2q^{2}\ln (\frac{r}{l})\right]
\label{BTZN2}
\end{equation}
where $N^{2}(r)$\ is known as the lapse function and $M$ and $q$ are the
mass and electric charge of the BTZ black hole, respectively.

\subsection{$(n+1)$-dimensional charged BTZ-like solutions\label{BTZlike}}
Here we want to obtain the $(n+1)$-dimensional static solutions of
Eqs. (\ref{GravEq}) and (\ref{MaxEq}). We assume that the metric
has the following form
\begin{equation}
ds^{2}=-F^{2}(r)dt^{2}+\frac{dr^{2}}{F^{2}(r)}+r^{2}\sum\limits_{i=1}^{n-1}d
\phi _{i}^{2},  \label{Metric}
\end{equation}
Again, we use the gauge potential like Eq. (\ref{Amu1}) in nonlinear Maxwell
equation (\ref{MaxEq}) for arbitrary $n$. We obtain surprisingly
\[
rh^{\prime \prime }(r)+h^{\prime }(r)=0,
\]
with the solution $h(r)=q\ln (\frac{r}{l})$ and so the electric
field is the same as $(2+1)$-dimensional BTZ solution (\ref{Ftr}).
It is notable that for higher dimensional linear Maxwell field
equation, the electric field depends on the dimensionality but in
our case (nonlinear Maxwell field), the electric field is
proportional to $r^{-1}$ for arbitrary dimensions.

In order to find the higher dimensional lapse function $F^{2}(r)$,
we should use the Eq. (\ref{GravEq} ). It is easy to show that the
solutions of all components of Eq. (\ref{GravEq}) can be written
as
\begin{equation}
F^{2}(r)=\frac{r^{2}}{l^{2}}-r^{2-n}\left[ {M+}2^{n/2}q^{n}\ln
(\frac{r}{l})\right] ,  \label{F(r)}
\end{equation}
where $M$ and $q$ are the integration constants which are related
to mass and charge parameters, respectively. As one can verify the
metric function $F^{2}(r)$ , presented here, differ from the
linear higher dimensional Reissner-Nordstr\"{o}m black hole
solutions; it is notable that the electric charge term in the
linear case is proportional to $r^{-2(n-2)}$, but in the presented
metric function, nonlinear case, this term is logarithmic. In the
$3$-dimensional limit ($n=2$), these solutions reduce to the known
BTZ solution (\ref{BTZN2}), and because of some similarities in
electromagnetic field and metric functions, hereafter we called
the higher dimensional solutions as BTZ-like solution.

\subsubsection{ Properties of the solutions}
In order to study the general structure of these spacetime, we
first investigate the effects of the nonlinearity on the
asymptotic behavior of the solutions. It is worthwhile to mention
that in arbitrary dimensions, the asymptotic dominant term of Eq.
(\ref{F(r)}) is first term and so BTZ-like solutions are
asymptotically AdS.

Second, we look for the essential singularities. After some algebraic
manipulation, one can show that for $(n+1)$-dimensional charged BTZ-like
solutions, the Ricci and Kretschmann scalars are
\begin{eqnarray}
R&=&-F^{\prime \prime }(r)-\frac{2(n-1)F^{\prime
}(r)}{r}-\frac{(n-1)(n-2)F(r)}{r^{2}} \nonumber \\
&=&\frac{-n(n-1)}{l^{2}}+\frac{2^{n/2}q^{n}}{r^{n}}, \label{R}
\end{eqnarray}
\begin{eqnarray}
R_{\mu \nu \rho \sigma }R^{\mu \nu \rho \sigma }&=&F^{\prime
\prime 2}(r)+\frac{2(n-1)F^{\prime 2}(r)}{r^{2}} \nonumber
\\ &&+\frac{2(n-1)(n-2)F^{2}(r)}{r^{4}}. \label{RR}
\end{eqnarray}
Also one can show that other curvature invariants (such as Ricci square,
Weyl square and so on) are functions of $F^{\prime \prime }$, $F^{\prime }/r$
and $F/r^{2}$ and therefore it is sufficient to study the Kretschmann scalar
for the investigation of the spacetime curvature singularity(ies).

Straightforward calculations show that presented Ricci and Kretschmann
scalars with metric function (\ref{F(r)}) diverges at $r=0$, is finite for $%
r\neq 0$ and for large values of $r$, we have
\begin{eqnarray}
\lim_{r\longrightarrow \infty }R &=&\frac{-n(n-1)}{l^{2}},  \label{AsympR} \\
\lim_{r\longrightarrow \infty }R_{\mu \nu \rho \sigma }R^{\mu \nu \rho
\sigma } &=&\frac{2n(n+1)}{l^{4}},  \label{AsympRR}
\end{eqnarray}
Equations (\ref{AsympR}) and (\ref{AsympRR}) confirm that the
asymptotic behavior of the solutions are AdS and Also we find that
there is a curvature timelike singularity located at $r=0$.

\begin{figure}[tbp]
\epsfxsize=8cm \centerline{\epsffile{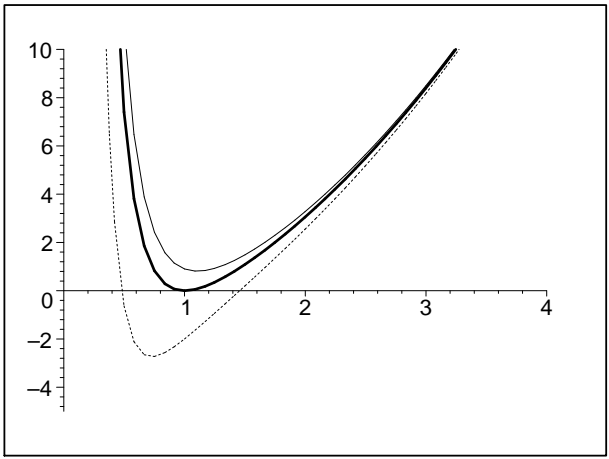}} \caption{The lapse
function, $F^2(r)$, of BTZ-like solutions versus $r$ for $n=4$,
$l=1$, $q=1$, and $M=0.1$ (no real solution or naked singularity:
continuous line), $M=1$ ($r_{-}=r_{+}=1$ or extreme black hole:
bold line), and $M=3$ ($r_{-}=0.48$, $r_{+}=1.46$ or two horizons:
dashed line).} \label{Fig1}
\end{figure}

It is easy to find that the presented black holes (BTZ and BTZ-like) have
two inner and outer horizons. For $(n+1)$-dimensional BTZ-like solutions
with arbitrary $n$, the horizons are located at%
\begin{eqnarray}
r_{-} &=&l\exp \left\{ -\frac{1}{n}L_{W}\left(
\frac{-nl^{n-2}e^{\left( \frac{-nM}{2^{n/2}q^{n}}\right)}}{
2^{n/2}q^{n}} \right) -\frac{M}{
2^{n/2}q^{n}}\right\} , \nonumber \\
r_{+} &=&l\exp \left\{ -\frac{1}{n}L_{W}\left(
-1,\frac{-nl^{n-2}e^{\left( \frac{-nM}{2^{n/2}q^{n}}\right)}}{
2^{n/2}q^{n}} \right) -\frac{M}{ 2^{n/2}q^{n}}\right\} , \nonumber
\end{eqnarray}
where the $L_{W}$ is $LambertW$ function satisfies
$LambertW(x)\exp \left[LambertW(x)\right]=x$ (for more details,
see \cite{Lambert} and figures \ref{Fig1}).
The event horizon is the hypersurface in which light can no longer
escape from the gravitational pull of a black hole. For
calculation of the event horizon, one can use the time dilation
interpretation (gravitational red shift) \cite{MTW}. It is
straightforward to show that the event horizon of the presented
BTZ-like solutions are located at the root(s) of $F^{2}(r)=0$.
Thus, the presented $r_{+}$ is the radius of the event horizon. In
addition, Fig.\ref{Fig1} shows that depending on the metric
parameters, the BTZ-like solutions may be interpreted as black
hole solutions with inner and outer horizons, an extreme black
hole or naked singularity.

\subsubsection{Conserved and thermodynamics quantities\label{Therm}}
In order to calculate the temperature, one may use
of the definition of surface gravity,
\[
T_{+}=\beta _{+}^{-1}=\frac{1}{2\pi }\sqrt{-\frac{1}{2}\left( \nabla _{\mu
}\chi _{\nu }\right) \left( \nabla ^{\mu }\chi ^{\nu }\right) }
\]
where $\chi =\partial /\partial t$ is the Killing vector. One obtains
\begin{equation}
T_{+}=\frac{r_{+}}{4\pi }\left( \frac{n}{l^{2}}-\frac{2^{n/2}q^{n}}{r_{+}^{n}%
}\right) .  \label{Temp}
\end{equation}
The electric potential $U$, measured at infinity with respect to the
horizon, is defined by \cite{Gub}
\begin{equation}
U=A_{\mu }\chi ^{\mu }\left\vert _{r\rightarrow \infty }-A_{\mu }\chi ^{\mu
}\right\vert _{r=r_{+}}=-q\ln (\frac{r_{+}}{l}).  \label{U}
\end{equation}

More than thirty years ago, Bekenstein argued that the entropy of a black
hole is a linear function of the area of its event horizon, which so-called
area law \cite{Bekenstein}. Since the area law of the entropy is universal,
and applies to all kinds of black holes in Einstein gravity \cite%
{Bekenstein,Hawking3}, therefore the entropy of the BTZ-like black holes is
equal to one-quarter of the area of the horizon, i.e.,
\begin{equation}
S=\frac{(2\pi r_{+})^{n-1}}{4}.  \label{Entropy}
\end{equation}
The electric charge of the black holes, $Q$, can be found by calculating the
flux of the electromagnetic field at infinity, yielding
\begin{equation}
Q=(2\pi )^{n-2}2^{(n-6)/2}nq^{n-1}.  \label{Charg}
\end{equation}
The present spacetime (\ref{Metric}), have boundaries with timelike ($\xi
=\partial /\partial t$) Killing vector field. It is straightforward to show
that for the quasi local mass we have
\begin{equation}
\mathcal{M}=\int_{\mathcal{B}}d^{n-1}\varphi \sqrt{\sigma }T_{ab}n^{a}\xi
^{b}=\frac{(2\pi )^{n-2}\left( n-1\right) M}{8},  \label{Mas}
\end{equation}
provided the hypersurface $\mathcal{B}$ contains the orbits of $\varsigma $.

Here, we check the first law of thermodynamics for our solutions. We obtain
the mass as a function of the extensive quantities $S$ and $Q$ (see appendix
for more details). One may then regard the parameters $S$, and $Q$ as a
complete set of extensive parameters for the mass $\mathcal{M}(S,Q)$ and
define the intensive parameters conjugate to them. These quantities are the
temperature and the electric potential%
\begin{equation}
T=\left( \frac{\partial \mathcal{M}}{\partial S}\right) _{Q},\ \ U=\left(
\frac{\partial \mathcal{M}}{\partial Q}\right) _{S}  \label{Dsmar}
\end{equation}
It is a matter of straightforward calculation to show that the intensive
quantities calculated by Eq. (\ref{Dsmar}) coincide with Eqs. (\ref{Temp})
and (\ref{U}) (see appendix for more details). Thus, these quantities
satisfy the first law of thermodynamics
\[
d\mathcal{M}=TdS+UdQ.
\]

\subsubsection{Stability in the canonical Ensemble}
Finally, we investigate the stability of charged charged BTZ-like
black hole solutions. The stability of a thermodynamic system with
respect to the small variations of the thermodynamic coordinates,
can be studied by the behavior of the energy $\mathcal{M}(S,Q)$
which should be a convex function of its extensive variable. In
the canonical ensemble, the charge is fixed parameter, and
therefore the positivity of the heat capacity
$C_{Q}=T_{+}/(\partial ^{2}\mathcal{M}/\partial S^{2})_{Q}$ is
sufficient to ensure the local stability. Considering the
appendix, we can find that $(\partial ^{2}\mathcal{M}/\partial
S^{2})_{Q}$ is
\begin{eqnarray}
\left( \frac{\partial ^{2}\mathcal{M}}{\partial S^{2}}\right) _{Q} &=&\frac{%
n }{32\pi ^{2}}\left[ \frac{4^{\frac{n}{n-1}}S^{\frac{2-n}{n-1}}}{(n-1)l^{2}}%
+ \frac{\left( \frac{2^{7}\pi ^{2}Q^{2}}{n^{2}}\right) ^{\frac{n}{2(n-1)}}}{
nS^{2}}\right]  \nonumber \\
&=&\frac{2r_{+}^{2(1-n)}}{(2\pi )^{n}}\left( \frac{nr_{+}^{n}}{(n-1)l^{2}}
+2^{n/2}q^{n}\right) .  \label{dMSS1}
\end{eqnarray}
It is clear that $(\partial ^{2}\mathcal{M}/\partial S^{2})_{Q}$ is positive
and so the heat capacity is always positive for $r\geq r_{ext}$, where the
temperature is positive (since for for $r\geq r_{ext}$, $F^{2}(r)$ is an
increasing function). Thus, the black hole is stable in the canonical
ensemble. It is notable that $r_{ext}$ is the root of temperature, given by
\begin{equation}
r_{ext}=\left( \frac{l^{2}}{n}\right) ^{1/n}\sqrt{2}q.  \label{rext}
\end{equation}

\subsection{CLOSING REMARKS}
One of the significant defects of higher dimensional
Reissner-Nordstr\"{o}m solutions is that in $3$-dimension, these
solutions does not reduce to BTZ black hole. In this paper, we
have been introduced an action of Einstein-nonlinear Maxwell
gravity in which its solutions are very like to BTZ black hole and
we have been called them as BTZ-like solutions. It is interesting
that, in all dimensions, the electric field of the solutions is
proportional to $r^{-1}$. Also, it is notable that the lapse
function of BTZ-like solutions is very similar to BTZ black hole
and its charge term is logarithmic. After these motivations, we
found that BTZ-like solutions have a curvature singularity with
two horizons, generally. Then, we calculated thermodynamics and
conserved quantities and showed that these quantities satisfied
the first law of thermodynamics. Finally, we calculated the heat
capacity of the BTZ-like black hole solutions and found that they
are positive for all the phase space, which means that the black
hole is stable for all the allowed values of the metric
parameters. This phase behavior is commensurate with the fact that
there is no Hawking--Page
transition for a black object whose horizon is diffeomorphic to $%
\mathbb{R}
^{p}$ and therefore the system is always in the high temperature
phase.

Finally, it is worthwhile to generalize our static solutions to
rotating BTZ-like black hole and it is left for the future.

\begin{acknowledgements}
This work has been supported financially by Research Institute for
Astronomy and Astrophysics of Maragha.
\end{acknowledgements}
\vspace{1cm}
\begin{center}
\textbf{APPENDIX}
\end{center}

In order to check the first law of thermodynamics, we rewrite the
quasilocal mass with respect to the entropy and charge.
Considering Eqs. (\ref{F(r)}), (\ref{Entropy}) and (\ref{Charg}),
we have
\begin{eqnarray*}
\mathcal{M}&=&\frac{\left( n-1\right) }{32\pi ^{2}}\left[
\frac{\left( 4S\right) ^{\frac{n}{n-1}}}{l^{2}}-\left(
\frac{2^{7}\pi ^{2}Q^{2}}{n^{2}}
\right) ^{\frac{n}{2(n-1)}}\ln \left( \frac{\left( 4S\right) ^{\frac{1}{%
n-1}}}{2\pi l}\right) \right] . \\
&& \\
d\mathcal{M} &=&\frac{n}{32\pi ^{2}}\left[ \frac{4^{\frac{n}{n-1}}S^{\frac{1%
}{n-1}}}{l^{2}}-\frac{\left( \frac{2^{7}\pi ^{2}Q^{2}}{n^{2}}\right) ^{\frac{%
n}{2(n-1)}}}{nS}\right] dS- \\
&&\frac{n}{32\pi ^{2}}\left( \frac{2^{7}\pi ^{2}Q^{\frac{2}{n}}}{n^{2}}%
\right) ^{\frac{n}{2(n-1)}}\ln \left( \frac{\left( 4S\right) ^{\frac{1}{n-1}}%
}{2\pi l}\right) dQ, \\
d\mathcal{M} &=&\frac{r_{+}}{4\pi }\left( \frac{n}{l^{2}}-\frac{2^{n/2}q^{n}%
}{r_{+}^{n}}\right) dS+\left( -q\ln (\frac{r_{+}}{l})\right) dQ.
\end{eqnarray*}
Then we should differentiate it to obtain
\begin{eqnarray*}
d\mathcal{M} &=&\frac{n}{32\pi ^{2}}\left[ \frac{4^{\frac{n}{n-1}}S^{\frac{1%
}{n-1}}}{l^{2}}-\frac{\left( \frac{2^{7}\pi ^{2}Q^{2}}{n^{2}}\right) ^{\frac{%
n}{2(n-1)}}}{nS}\right] dS- \\
&&\frac{n}{32\pi ^{2}}\left( \frac{2^{7}\pi ^{2}Q^{\frac{2}{n}}}{n^{2}}%
\right) ^{\frac{n}{2(n-1)}}\ln \left( \frac{\left( 4S\right) ^{\frac{1}{n-1}}%
}{2\pi l}\right) dQ.
\end{eqnarray*}
Here, we should replace $Q$ and $S$ from Eqs. (\ref{Entropy}) and (\ref%
{Charg}), and rewrite $d\mathcal{M}$
\[
d\mathcal{M}=\frac{r_{+}}{4\pi }\left( \frac{n}{l^{2}}-\frac{2^{n/2}q^{n}}{%
r_{+}^{n}}\right) dS+\left( -q\ln (\frac{r_{+}}{l})\right) dQ.
\]
It is completely clear that the coefficients of $dS$ and $dQ$ are
temperature $T$ and electric potential $U$, respectively and so we
have
\[
d\mathcal{M}=TdS+UdQ.
\]

\end{document}